\documentstyle[12pt,aaspp4]{article}

\received{}
\revised{}
\accepted{}
\cpright{}{1996}

\journalid{}{}
\articleid{}{}
\paperid{}

\cpright{}{}
\ccc{}

\begin{document}

\newcommand{\kms}{km s$^{-1}$}

\title{G55.0+0.3: A Highly Evolved Supernova Remnant}
\author{B.~C.~Matthews\altaffilmark{1,2}}
\author{B.~J.~Wallace\altaffilmark{3}} 
\author{A.~R.~ Taylor\altaffilmark{1}} 

\altaffiltext{1}{Dept. Physics \& Astronomy, University of Calgary,
Calgary, AB, Canada, T2N 1N4} 
\altaffiltext{2}{Dept. Physics \& Astronomy, McMaster University, 
Hamilton, ON, Canada, L8S 4M1}
\altaffiltext{3}{Dominion Radio Astrophysical Observatory, P.~O. 
Box 248, Penticton, BC, Canada, V2A 6K3}

\begin{abstract}

In order to determine the nature of an object identified as a
potential supernova remnant (SNR) in a 327 MHz survey of the Galactic
plane, continuum observations have been taken at 1420 MHz with
resolution identical to that of the survey.  Additionally, atomic
hydrogen (H{\sc i}) spectral line data have been taken in order to
determine the distance to the object.
 
Multi-frequency analysis shows that a shell feature in the southern
part of this object is non-thermal and confirms earlier studies that
identified the northern portion as an H{\sc ii} region.  The
non-thermal nature of the shell and the absence of infrared flux
density confirms the speculation of Taylor, Wallace, \& Goss (1992)
that this object, G55.0+0.3, is a SNR.  From analysis of H{\sc i}
data, the SNR's estimated kinematic distance is 14 kpc, yielding a
radius of approximately 70 pc, making it one of the largest known
SNRs.  The age estimate on the order of one million years exceeds the
conventional limits on observable SNR lifetimes in the literature by a
factor of five, implying that the radiative lifetimes of SNRs could be
much longer than previously suggested.

There is also evidence that the remnant may be associated with the nearby
pulsar J1932+2020.  This proposed SNR/pulsar association is older
than any previously documented by an order of magnitude.  If valid, it
suggests that searches for associations should not be restricted to
the regions about young pulsars.

\end{abstract}

\keywords{supernova remnants --- ISM: individual (G55.0+0.3,
G55.2+0.5, G55.6+0.7 --- pulsars: individual (J1932+2020) --- radio
continuum: ISM}

\section{Introduction}

There are 215 supernova remnants (SNRs) presently known in the Milky Way
(\markcite{gre96}Green 1996).  Estimates of the fraction which could
have been created by Type II supernovae (the core-collapse of massive
stars) range from 75\% (\markcite{van87}van den Bergh, McClure, \&
Evans 1987) to 85\% (\markcite{tam94}Tammann, L\"{o}ffler, \&
Schr\"{o}der 1994).  Such collapses are also believed to form neutron
stars, sometimes observable as pulsars.  It has thus been expected that some
large fraction of detectable SNRs should be associated with young
pulsars.  In fact, only a small number of observed associations exist;
\markcite{car95}Caraveo (1995) summarizes 13 such associations, while
Gaensler \& Johnston (\markcite{gae95c}1995a) note that a total of 18
have been postulated.

\markcite{gae95b}Gaensler \& Johnston (1995b) discuss reported
SNR/pulsar associations and suggest from a statistical analysis that
the expected number of observable associations is actually less than
the reported number.  The number of expected associations is strongly
dependent on the observable lifetimes of SNRs.  Currently, the oldest
age estimates are on the order of 100,000 yr (e.g.~CTB 80,
\markcite{shu89}Shull et al.~1989).  Old radio remnants are likely to
be physically large and have low surface brightnesses.  Such objects
may be difficult to separate from the Galactic background and are thus
likely to be under-represented in current catalogues of SNRs.

Several very old SNRs may have already been detected, but lack of
reliable distance estimates prevents the determination of either their
physical sizes or their ages.  Thus, our knowledge of SNR lifetimes is
constrained more by theory than by observation.  The detection of
large and faint remnants can best be accomplished in high sensitivity,
wide-field surveys of the Galactic plane. One such survey, conducted
with the Westerbork Synthesis Radio Telescope (WSRT) at 327 MHz was
sensitive to structures up to 45\arcmin \ in size
(\markcite{tay96}Taylor et al.~1996) and detected a number of new SNR
candidates (\markcite{tay92}Taylor et al.~1992).  One of these
candidates, G55.2+0.5, is a shell-like object coincident in position
with a pulsar of spindown age $1.1 \times 10^6$ years, PSR J1932+2020.
An association between the SNR and this pulsar, discovered in 1974
(\markcite{hul74}Hulse \& Taylor 1974, \markcite{hul75}1975), had been
suggested previously by \markcite{cro74}Crovisier (1974), but
\markcite{rei86}Reich et al.~(1986) suggested instead that the
emission in this region is produced by an H{\sc ii} region, not a SNR.

This paper's purpose is two-fold: to examine whether G55.2+0.5 is or
contains a SNR; and to examine the possible association with PSR
J1932+2020.  To these ends, we have obtained new continuum
observations of the region around G55.2+0.5 at 1420 and 408 MHz, as
well as atomic hydrogen (H{\sc i}) line observations.  These
observations are discussed in $\S$\ref{sec-obs}.  In
$\S$\ref{sec-res}, we present the new data and compare them to data at
a number of other wavelengths.  The results of this analysis are
discussed in $\S$\ref{sec-dis} and summarized in $\S$\ref{sec-conc}.

\section{The Observations}
\label{sec-obs}

The observations for this investigation were carried out at the
Dominion Radio Astrophysical Observatory (DRAO) in Penticton, British
Columbia, Canada.  The observatory is comprised of a synthesis
telescope of seven antennae, each 9 m in diameter, and a 26 m
single-dish antenna. The synthesis telescope is laid out on an
east-west track.  The second, third and fourth antennae are moveable,
although they maintain a fixed spacing with respect to one another.
These three antennae are moved daily over a period of twelve days so
that the entire range of possible baselines from 12.9 m to 604.3 m are
sampled at intervals of 4.3 m.

The DRAO synthesis telescope observes simulateously in continuum at
408 and 1420 MHz and on the spectral line of H{\sc i} at 1420 MHz
($\lambda 21$ cm).  Short-spacings H{\sc i} line data (to fill in
missing baselines shorter than 12.9 m) are obtained with the 26 m
dish. More information on the DRAO synthesis telescope can be found in
papers by \markcite{rog73}Roger et al.~(1973) and
\markcite{vei85}Veidt et al.~(1985).


The data were taken during the period from 6 August, 1992 to 23
September, 1992; one baseline was re-observed on 7 January, 1993.  The
field center is at right ascension $19^{\rm h} 32^{\rm m} 0^{\rm s}$
and declination $20^{\rm d}$ 6\arcmin \ 0\arcsec \ in J2000 coordinates.
This position lies only 0.5\arcdeg \ above the Galactic plane.

These data were among the first obtained after the DRAO upgrade from 4
to 7 antennae.  The 1420 MHz continuum image contains an artifact at
its centre which resulted from a correlator error that was present for
a short period after the upgrade but was quickly fixed.  Had the
artifact been beam-shaped, we could have easily modeled and removed
it.  Despite initial appearances, however, it is not a signal which we
would have expected from a point source at the centre
(i.e.~beam-shaped), and alternative methods were required to address
the problem. To minimize the effects of the error, archival data from
the same epoch were obtained from a high-declination field which did
not have a source at the map centre.  From these data, a model of the
artifact containing minimal confusion with background sources was
extracted.  This artifact was corrected for the difference in
north-south resolution of the interferometer and subtracted from the
$u-v$ data of our field.  A difference in the intensities of the two
artifacts, the cause of which is not clear, required that only a
fraction of the high-declination artifact be subtracted; the decision
as to which fraction yielded the most satisfactory results was made by
eye.  The result of the subtraction was cosmetic improvement, but
values of the affected pixels at the centre of the map remain
unreliable.  Fortunately, this region is not near the region of
interest, which actually surrounds, but does not include, the centre
of the map.

The continuum DRAO fields were self-calibrated and CLEANed using DRAO
software.  As with all interferometers, large spatial scale emission
cannot be detected due to missing short-spacings.  In the case of the
DRAO interferometer, features larger than about one degree in size
cannot be detected at 1420 MHz by the array (three degrees at 408
MHz).  Single-dish data must be obtained to account for emission from
large-scale sources.  For this purpose, data from the 21 cm Effelsberg
survey of \markcite{rei90}Reich, Reich, \& Fuerst (1990) and the 408
MHz all-sky survey of \markcite{has83}Haslam et al.~(1983) were used.

The specifications for the DRAO continuum data are given in Table
\ref{DRAOspecs-st}.  Estimates of rms noise in the images were obtained by
direct measurement of map fluctuations in a region of relatively low
emission and are included in the table.

The DRAO synthesis telescope obtains H{\sc i} line data at both left-
and right-handed polarizations.  The two polarizations were averaged
together before mapping.  The specifications of these data are
summarized in Table \ref{DRAOspecs-HIst}.  The level of emission in
the H{\sc i} data cube is sufficiently low that no self-calibration or
CLEANing is possible.  The continuum contribution to the H{\sc i} cube
was removed by subtracting a continuum map made from emission-free
maps at either end of the H{\sc i} data cube.  


To obtain spectral information about the continuum emission, we have
also used the 327 MHz data of \markcite{tay96}Taylor et al.~(1996) and
the 2695 MHz data of \markcite{rei84}Reich et al.~(1984).  In
addition, we have obtained HiRes IRAS data (\markcite{aum90}Aumann,
Fowler, \& Melnyk 1990; \markcite{cao96}Cao et al.~1996;
\markcite{fic96}Fich \& Terebey 1996) for this region at 60 and 100
microns.  The details of these datasets are summarized in Table
\ref{otherfreqtable}.  This table contains the resolution of the data
sets (in the case of the IRAS data, these are the HiRes values);
the position angle (PA) east of north of the major axis of the beam;
and the rms noise in the maps.

\section{Results and Analysis}
\label{sec-res}

\subsection{Radio Spectral Indices}

The 327 and 1420 MHz data are presented in Figure \ref{cont-images}.
The images at these two frequencies have almost identical angular
resolution.  The 327 MHz image is purely interferometric and therefore
contains no large scale smooth structures (i.e.~greater than 45\arcmin
\ in angular size).  The region of interest lies primarily to the
north and south of the centre of the image and has three components --
two bright compact regions of emission to the north and a partial
shell structure to the south.  \markcite{tay92}Taylor et al.~(1992)
suggested these features may represent one object, a shell about
0.5\arcdeg \ in diameter, designated by its central position of
G55.2+0.5, as shown in Figure \ref{cont-images}.

The centres of the northern structure and southern shell have
coordinates G55.6+0.7 and G55.0+0.3 respectively.  G55.6+0.7 contains
two compact features superimposed on more extended emission. The
position of the pulsar J1932+2020, marked by a cross on the figures,
is within the east component of G55.6+0.7.  The properties of the
pulsar are summarized in Table \ref{psr-parameters} based on
measurements by \markcite{lor95}Lorimer et al.~(1995).  The
uncertainties refer to the last significant digits in the
measurements.

It is noteworthy that the coordinates G55.6+0.7 are those which were
assigned to the radio source identified first as a potential SNR by
\markcite{cro74}Crovisier (1974) and later as a probable H{\sc ii}
region by \markcite{rei86}Reich et al.~(1986).
Aside from the features identified as components of a possible remnant
by \markcite{tay92}Taylor et al.~(1992), a portion of the SNR HC 40
(G55.4$-$0.3) is visible as a saturated bright arc to the south-east
in Figure \ref{cont-images}.  Other extended sources are visible as
well; however, they are of smaller angular extent than G55.2+0.5 and
HC 40.

Although they have poorer resolution than the 327 and 1420 MHz maps,
maps of the region at 408 and 2695 MHz also reveal strong emission
coincident with the features in the 327 and 1420 MHz maps.  However,
the individual components of G55.6+0.7 are not distinguishable at
these frequencies.  

The integrated flux densities of G55.6+0.7 and G55.0+0.3 were
determined by summing the map intensities over the sources.  The
region of integration was set by enclosing each source in a polygon
shape constructed on the image.  The integrated flux density is the
sum of the intensities within the polygon above a background plane,
which is defined by fitting to the intensities at the vertices of the
polygon.  Before carrying out this procedure, the contribution from
``point'' (i.e.~unresolved) sources within the polygon regions was
removed.  These sources are typically extragalactic and are removed so
that they do not contribute to the measure of flux density from the
extended, Galactic source.

Point source removal was performed on the 327 and 1420 MHz maps using
Gaussian fitting.  Gaussian fits could not be performed directly on
the 408 and 2695 MHz images due to poor resolution.  The flux
densities obtained at 327 and 1420 MHz were used to extrapolate the
408 and 2695 MHz point source flux densities.  These were subtracted
from their integrated flux densities to remove the point source
effects.

The presence of the pulsar added an additional difficulty to the
integration of flux density over G55.6+0.7 and G55.2+0.5.  The
integrated flux density of pulsars may vary on long time scales, and
there has been significant variation in the flux density measurements
of this pulsar, with data from different epochs and with telescopes of
different beamsizes.  We have extrapolated all flux densities using
the flux densities and spectral index of \markcite{lor95}Lorimer et
al.~(1995).  


At 327 MHz, the pulsar is visible in the image as an eastward
extension of G55.6+0.7.  The flux density at that point in the image
is 40.5 mJy.  The extrapolated flux density of the pulsar at 327 MHz
from \markcite{lor95}Lorimer et al.~(1995) is $49.6\pm 12$ mJy.  For
the purposes of removal of the pulsar's flux density, the most
conservative estimate was taken; all the flux density observed at the
pulsar's position was attributed to the pulsar itself.  The
contribution from the pulsar was far less significant at 1420 MHz,
with an extrapolated flux density of $1.18\pm 0.4$ mJy compared to an
observed flux density of 211 mJy at the pulsar position.  Thus, at
this frequency, the pulsar's emission accounts for less than 1\% of
the observed flux density.

The polygons were created to match the regions of emission at 327 and
1420 MHz.  Several polygons with approximately the same shape but
different vertices were used to obtain independent estimates of the
background, which is the greatest source of uncertainty in the
determination of flux densities. The sums over polygons above
background can be as little as 5\% of the total flux densities.  The
backgrounds of each polygon were fit using a least-squares method to
the edges of the region enclosed by the polygons and then subtracted.
The resulting flux densities are listed in Table \ref{intfluxtable}.
The uncertainties are the standard deviations of the polygon means.

Using the integrated flux densities, the average spectral indices of
the various features were determined.  For each region, this was done
by performing a weighted least-squares fit to the flux densities at
each frequency.  The resultant values for the radio spectral index are
given in Table \ref{intalphatable}. The radio continuum spectra of
G55.6+0.7 and G55.0+0.3 are plotted in Figure \ref{spectra}. The
integrated spectrum of G55.0+0.3 indicates non-thermal emission since
$\alpha = -0.53\pm 0.26$ where $S_\nu \propto \nu^\alpha$.  The
average spectral index for known shell-type remnants is $-0.47 \pm
0.16$ based on values from \markcite{gre96}Green (1996).  The northern
source G55.6+0.7 could be indicative of thermal emission, since
$\alpha = 0.01\pm 0.10$.

The integrated flux densities provide a measure of the mean spectral
index over each region.  To look for a change in the spectral index
within each region, we constructed spectral index maps from the 327
and 1420 MHz brightness temperature ($T_b$) images.  Before constructing a 
spectral index map, the local background flux densities were removed
by fitting independently to local regions around G55.6+0.7 and
G55.0+0.3, and the pulsar's flux density was subtracted from the
images.

Spectral index ratio maps are subject to large uncertainties in
regions where the denominator image is close to zero.  Thus, spectral
indices were calculated only for those pixels with brightness
temperatures greater than 12 K($T_b$) above background at 327
MHz. This is the 3$\sigma$ threshold, based on the measured noise in
the 327 MHz map (see Table \ref{otherfreqtable}).

Uncertainties in the spectral index map from the noise present in the
327 and 1420 MHz maps are of order 0.1.  After the removal of the
pulsar, there was no significant variation in the spectral index of
G55.6+0.7 and no evidence of non-thermal emission in any part of that
region.

The spectral index map of G55.0+0.3 is shown in Figure \ref{mouth}.
The quantity plotted is $2-\alpha$.  Significant, systematic
variations in the brightness temperature spectral index appear within
this structure. Those regions which appear black ($\ge 2.2$) are
unambiguously non-thermal (the threshold value is 2.1).  The white
region of the map contains pixels for which the spectral index is not
defined due to the low 327 MHz flux density.  

Although most of the shell clearly shows non-thermal spectral indices,
there is a large section in the southwest portion of the G55.0+0.3
which shows a flatter spectrum that could indicate either thermal or
non-thermal emission.  The presence of a thermal object in the
foreground of the SNR could produce such a spectral index, or there
may genuinely be a weaker shock at that position, decreasing the
amount of non-thermal emission so the thermal background radiation is
more prominent.  The minimum value of the spectral index is $\alpha =
-1.47$ at the southeast outer edge, while the maximum value is $\alpha
= 0.34$ within the region to the southwest.

The spectral index analyses support both the suggestion of
\markcite{tay92}Taylor et al.~(1992) that the region G55.2+0.5
contains a SNR (although it does not include the northern region) and
the suggestion by \markcite{rei86}Reich et al.~(1986) that the
northern collection of objects, G55.6+0.7, is an H{\sc ii} region.

\subsection{IRAS Data}

Figure \ref{iras-radio} shows the HiRes 60 $\mu$m image of this region
with 327 MHz continuum contours superimposed.  No emission is present
that is correlated to G55.0+0.3.  However, there is heightened
emission coincident with the northern source, G55.6+0.7.

Although a lack of 60 $\mu$m emission alone is insufficient to
discriminate between objects of potentially similar morphology such as
H{\sc ii} regions and SNRs, IRAS data in conjunction with radio data
may be used to distinguish between objects of similar morphology in
the two regimes (\markcite{are89}Arendt 1989; \markcite{sak92}Saken,
Fesen, \& Shull 1992). \markcite{tay92}Taylor et al.~(1992) denoted
values of the 60 $\mu$m to 327 MHz flux density ratios below 200 to be
indicative of non-thermal sources.  Table \ref{intalphatable}
summarizes the ratio for each region. These results support the
conclusions of the radio analysis that G55.0+0.3 is an SNR and that
G55.6+0.7 is an H{\sc ii} region.

\markcite{shu89}Shull et al.~(1989) note that some old remnants may be
weak in radio, optical and X-ray emissions due to their low shock
velocities, yet be extremely bright in the infrared.  However, only
one third of radio SNRs have been detected at infrared wavelengths,
and faint radio remnants (e.g.~G152.2$-$1.2, G291.9+5.5) are
conventionally among those which have no detectable emission in the
infrared (\markcite{are89}Arendt 1989).  Thus, absence of IRAS
emission might be expected for an evolved or faint SNR.

\subsection{H{\sc i} Line Data}

The H{\sc i} line data were used to search for evidence of atomic
hydrogen associated with the radio continuum shell.  The only obvious
correlation was noted over nine velocity channels ranging from $-39.6$
to $-52.8$ \kms.  Over this range, a void which matches the size and
shape of the outer edge of the continuum emission shell is present.
No evidence was found for high-velocity H{\sc i} related to the SNR
in the ``empty'' channels of the H{\sc i} cube.

Figure \ref{HI89-97fig} shows the velocity channels over which the
void is present as well as additional channels on either end.  For
ease of presentation, a mean intensity bias has been subtracted from
each channel.  The correlation is strongest for the central channel,
as would be expected for a symmetric expanding cavity and supports
the notion that the void is related to the SNR.  A continuum overlay
on the central channel ($-46.2$ \kms) is shown in Figure
\ref{channel93}.

The application of a flat Galactic rotation model with $R_\odot = 8.5$
kpc and $V_\odot = 220$ \kms \ yields a conversion of the LSR velocity
to a kinematic distance of 14 kpc.  The uncertainty associated with
kinematic velocities due to the assumption of circular rotation of the
Galaxy is on the order of 10\% (\markcite{bur88}Burton, 1988).  The
distance of G55.0+0.3 from the Galactic centre would therefore be
$\sim 11$ kpc, and its height above the Galactic plane would be $\sim
73$ pc.

We superimposed a circle of radius 17.5\arcmin \ to the shell of
continuum emission, as shown in contour in Figure \ref{channel93}. At
14 kpc, this angular size implies a physical radius of $\sim 70$ pc.
This size is large, but not unprecedented.  Table \ref{sizes}
summarizes the sizes of some of the largest SNRs for which distance
estimates exist.  The average physical radii ($R_{avg}$) are
calculated for the lower distance limits ($d_l$) in the literature,
which yield the lower limits on physical dimension.

\section{Discussion} 
\label{sec-dis}

In summary, our analysis of the radio and 60 $\mu$m emission indicates
that the half-shell of emission designated G55.0+0.3 is a supernova
remnant.  If it lies at a distance of 14 kpc as suggested by the
association with the H{\sc i} void at $v_{LSR} = -46.2$ \kms, its
linear diameter is approximately 140 pc.  

The 1 GHz surface brightness of the new remnant G55.0+0.3 is $(7.4\pm
1.7) \times 10^{-23}$ W m$^{-2}$ Hz$^{-1}$ sr$^{-1}$, based on the
angular size of $8.1 \times 10^{-5}$ sr for the circular fit shown in
Figure \ref{channel93}.  Only one other SNR (G156.2+5.7) has a
comparably low value of $6.2\times 10^{-23}$ W m$^{-2}$ Hz$^{-1}$
sr$^{-1}$.  Based on the solid angle of the continuum shell alone of
$2.9\times 10^{-5}$ sr, the surface brightness of the visible portion
of the remnant is $(2.1\pm 0.5) \times 10^{-22}$ W m$^{-2}$ Hz$^{-1}$
sr$^{-1}$.  Thus, the remnant is among the faintest 10\% of all
remnants.  This surface brightness, combined with its large angular
and linear dimensions, suggests that it is highly evolved.  In the
next section, we attempt to place constraints on the age and
evolutionary phase of G55.0+0.3.

\subsection{Age via Evolutionary Models}

As they expand, shell-type SNRs pass through four reasonably distinct
phases as they merge into the interstellar medium (ISM).  The four
phase evolution, notwithstanding refinements, was first summarized by
Woltjer (1972).  The first phase, referred to by some as ``free
expansion'', is characterized by a double-shock structure
(\markcite{mck74}McKee 1974).  After the blast wave has swept through
a significant amount of the ISM, the SNR enters the Sedov-Taylor
phase, modeled as a point-blast into a uniform medium
\markcite{sed59}(Sedov 1959) and characterized by a $r^{2/5}$
expansion in time.  This is followed by radiative expansion (when
radiative cooling by metals causes a dense shell to form) and,
finally, the merging of the remnant with the surrounding ISM.

The large physical size, low surface brightness and clumpy morphology
of this remnant argue that it is in the late Sedov-Taylor or radiative
phase of its expansion.  After \markcite{shu89}Shull et al.~(1989), we
use
\begin{equation}R_s(t)=31.5\ E_{51}^{0.2} n_0^{-0.2} t_5^{0.4}
\ \ \ \rm{(pc)}
\label{Sedov-radius}
\end{equation}

\noindent for the radius of a remnant in the Sedov-Taylor expansion
stage, where $E_{51}$ is the initial kinetic energy of the ejecta in
units of $10^{51}$ erg, $n_0$ is the ambient interstellar hydrogen
density in cm$^{-3}$ (assumed to be constant), and $t_5$ is the SNR's
age in units of $10^5$ yr.  

For times much greater than the onset of the pressure-driven snowplow,
or radiative, phase, the model of \markcite{cio88}Cioffi, McKee, \&
Bertschinger (1988) predicts the radius of the shell to be given by
\begin{equation}R_s(t)\approx 27.9\ E_{51}^{0.221} n_0^{-0.257}
\xi_m^{-0.0357} t_5^{0.3} \ \ \ \rm{(pc)}
\label{Snow-radius}
\end{equation}

\noindent which takes into account the the dynamics of the shell
evolution and a radiative cooling function,
\begin{equation}\Lambda(T)=(1.6\times10^{-19})\xi_mT^{-0.5} 
\ \ {\rm (erg \ cm^{-3} \ s^{-1})} 
\label{cool-fn}
\end{equation}

\noindent where $\xi_m$ is metallicity (equal to unity for solar
abundances) and $T$ is temperature.  We have used $\xi_m = 1$
throughout our analysis.  The form and constants of equations
(\ref{Sedov-radius}) and (\ref{Snow-radius}) are obviously dependent
on assumptions of individual models.  The above equations were
generated for conditions of neglible presssure, spherical expansion, a
homogeneous and uniform ISM and no thermal conduction (for more
details, see Cioffi et al., 1988).

For G55.0+0.3, setting $R_s = 70$ pc in equations (\ref{Sedov-radius})
and (\ref{Snow-radius}) yields expressions for the ages in the
Sedov-Taylor and radiative phases to be:
\begin{equation}t_5^{II} = 8 \ E_{51}^{-0.5} n_0^{0.5} \ \ \ \ 
(10^5 \ \rm{yr})
\label{adi-age}
\end{equation}

\noindent and 
\begin{equation}t_5^{III} = 24 \ E_{51}^{-0.737} n_0^{0.857} 
\xi_m^{0.119} \ \ \ (10^5 \ \rm{yr}).
\label{rad-age}
\end{equation}

\noindent where superscripts II and III refer to the Sedov-Taylor and
radiative phases, respectively.

The evolutionary model can be used to set reasonable limits on how old
the remnant will be at the transition times between phases.  These
calculations can be used to establish, for a given distance, the
evolutionary phase of the SNR.

The transition from Sedov-Taylor to radiative expansion occurs at
\begin{equation}t_{rad}=(1.33\times 10^4)\ E_{51}^{3/14} n_0^{-4/7} 
\xi_m^{-5/14} \ \ \ \rm{(yr)}
\label{t-rad}
\end{equation}

\noindent (\markcite{cio88}Cioffi et al.~1988) where the radiative
cooling function given by equation (\ref{cool-fn}) has been used.
Equating equations (\ref{adi-age}) and (\ref{t-rad}) yields the
condition required for the remnant to be younger than $t_{rad}$.  This
condition can be written as
\begin{equation}E_{51} \ge 300 \ n_0^{1.5} 
\label{equate}
\end{equation}

\noindent for density in units of cm$^{-3}$.  If the remnant is in the
Sedov-Taylor phase, then the inequality holds.  Observations of
extragalactic supernovae (SNe) give $E_{51} = 0.5$ to 2
(\markcite{fil94}Wheeler \& Filippenko 1994; \markcite{arn89}Arnett et
al.~1989).  The typical value for Type II SNe is $10^{51}$ erg
(\markcite{che77}Chevalier 1977).  We adopt an upper limit of 2 for
$E_{51}$ and thus obtain an upper limit on the ambient density, $n_0$,
for the inequality to hold.  Substitution into inequality
(\ref{equate}) gives an upper limit on the density of 0.04 cm$^{-3}$.
Unless $n_0 < 0.04$, the remnant is in the radiative phase.

Unfortunately, once a SNR enters Phase II, the information about the
initial energy of the SN is not retained.  However, estimation of the
ambient density of the material with which the remnant is currently
interacting can be made using the H{\sc i} line data.

If the void in H{\sc i} is taken to be the cavity in the ISM swept up
by the remnant, we can use its diameter to estimate the pathlength
through the H{\sc i} which is occupied by the remnant, under the
assumption that it has evolved in a roughly spherically symmetric
fashion south of the blast position (i.e.~it has formed a recognizable
half-shell).  The velocity width of the void therefore represents the
velocity width of the H{\sc i} evacuated by the remnant.  In this
case, the gas adjacent to the void within the same velocity interval
is the unswept-up ISM occupying the same physical pathlength as the
depth of the void.  If the void is as deep as it is wide, this depth
is 140 pc.

The density of the gas within this pathlength can be obtained from the
column density over this velocity interval.  It is obtained by
integrating the brightness temperature (a sum of the velocity channels
over which the void is evident) according to the expression:
\begin{equation}n_0 = \frac{1.822 \times 10^{18}}{l} \Delta v 
\sum T_B(v)
\label{eqn-density}
\end{equation}

\noindent where $l$ is the length of the void along the line of sight
in centimetres and $\Delta v$ is the channel width in \kms \
(\markcite{roh86}Rohlfs 1986).  Figure \ref{density} is a map of $n_0$
in the region of the remnant calculated from the nine velocity
channels from $-39.6$ to $-52.8$ \kms.  The values are of order unity
everywhere adjacent to the shell.  A large, low density region appears
to the north.  The low density of the ISM in this direction may
explain the absence of a non-thermal shell there.  Equations
(\ref{adi-age}) and (\ref{rad-age}) both support faster expansions
into regions of lower density, so there may be no evidence of a
remnant remaining to the north.

The presence of turbulent motions in the ISM actually prevents the
measurement of emission uniquely from regions as small as 140
pc. These turbulent motions will contribute emission from gas outside
this pathlength interval.  At the distance of the void of 14 kpc, the
typical turbulent velocity of 8 \kms\ (\markcite{spi78}Spitzer 1978)
corresponds to a distance along the rotation curve of almost 2 kpc.
Therefore, emission from gas displaced by up to $\pm 900$ pc could
spill over into the velocity interval of the void.

We have dealt with this excess of gas by presuming that the void is
completely devoid of H{\sc i}.  Thus, the density estimate within the
void is in effect a measure of the ``background'' contributions of the
turbulent gas of the ISM.  We also assume uniformity across our region
of interest and thus simply subtract the ``background'' from the
density estimates outside the void.

Taking into account the subtraction of the density estimate within,
but near the edge of, the void ($\sim 1.6 \pm 0.1$ cm$^{-3}$) and the
value just outside the void ($\sim 2.4 \pm 0.1$ cm$^{-3}$), this
yields an ambient density estimate of $0.8 \pm 0.2$ cm$^{-3}$ for the
pre-swept-up gas, at least in the regions adjacent to the continuum
half-shell.  Over this half sphere, this density implies a swept-up
mass of $\sim 10,000 M_\odot$ under the gross assumption (but one
which we made above to estimate the contribution of turbulent motions
in the ISM) that this cavity was swept clean by the expanding blast
wave.  This estimate of the swept-up mass is clearly an upper limit
(see below).


In summary, the critical assumptions in this calculation are: that the
cavity is as deep as it is wide (for the range of angles which the
shell subtends); that the cavity is in effect devoid of H{\sc i}; and
that the contributions from the background ISM are uniform across and
just outside the remnant.  We have no way of estimating the density of
the ISM in the past; thus, we also assume to first-order that the
density of the ambient medium into which the remnant is currently
expanding is the same as that of the material it has already swept up.
Measurements of molecular material would be useful for obtaining
further information about the region into which G55.0+0.3 is evolving.

A density of less than 0.04 cm$^{-3}$ for the material {\it adjacent}
to the shell is essentially ruled out by this analysis. Therefore, we
conclude that the remnant is in the radiative phase, since the
inequality of equation (\ref{equate}) does not hold.  Additionally,
unless our H{\sc i} density estimate is incorrect by a factor greater
than 1.5 orders of magnitude, the order of the resultant age will
remain unchanged.

We have defined the time of the merger of the remnant with the ISM
(dissipation) to be that at which the shock has slowed to 8 \kms, the
speed of macroscopic turbulence in the ISM (\markcite{spi78}Spitzer
1978).  Differentiating equation (\ref{Snow-radius}) with respect to
time and setting the result equal to the turbulence speed yields
\begin{equation}t_{merger}=(2.9 \times 10^6)\ E_{51}^{0.316} n_0^{-0.367} 
m^{-1.43}\ \ \ \rm{(yr)}
\label{mergertime}
\end{equation}

\noindent where $m$ is the local sound speed in units of 8 \kms. 

If the density is indeed $0.8\pm 0.2$ cm$^{-3}$, then the remnant has
an age of $(1.9\pm 0.4) \times 10^6$ yr, taking $E_{51} = 1$.  The
same model gives a merger time for the remnant of $3.1_{+0.3}^{-0.2}
\times 10^6$ yr.  A decrease in SN energy of $\sim 35$\% produces
radiative ages in excess of merger times while a corresponding
increase inflates the merger time, but decreases the radiative age to
just over one million years.

The ultimate ages obtained for the remnant depend on where one
estimates the remnant boundary to be.  However, a decrease in the
physical size in excess of 20\% is required to alter the radiative age
to below one million years.  This radius is well within the boundaries
of the continuum shell.

Many assumptions were made to derive the density above, as we have
already stated.  Several of these have predictable consequences for
the derived density.  For instance, the assumption that the remnant
has expanded into a region of uniform density is clearly a
non-physical one, especially given its physical size.  It is possible
that the size of the void in the H{\sc i} associated with the SNR owes
its existence only partially or not at all to the actions of the SNR.

It is well-known that massive stars have powerful winds and are
capable of sweeping out giant ``bubbles'' in the ISM over their
lifetimes.  Under such circumstances, the SNR formed by the final
explosion of the star initially expands into a region of depleted
ambient density.  If this has indeed been the case for the remnant
G55.0+0.3, then the age estimate calculated above is incomplete, since
the remnant would have initially expanded into a region of
circumstellar material of relatively low density.
\markcite{mck84}McKee, van Buren, \& Lazareff (1984) conclude that a
massive star which ends its life in a Type II SN will interact with
circumstellar material to a radius of about 20 pc.  Due to the low
density in this regime, the total mass within this radius is
comparable to the total mass of the stellar ejecta.  Thus, the remnant
would expand rapidly to this radius.  The radius of the remnant
G55.0+0.3 exceeds this value by greater than three times, implying
that it should have evolved well past Phase I, even without taking
into account its advanced morphology.

However, larger voids in the ISM can be created by clusters of massive
stars during their lifetimes, or by the SNRs they spawn after death.
If such a structure was created at the location of the SN which
created G55.0+0.3, then the size of the currently visible void is the
same as it was when the SN occurred.  This scenario would invalidate
all the previous dynamical calculations, since our estimate of the
upper limit on the age assumes that the SNR has been interacting with
an ISM of density 0.8 cm$^{-3}$ for some large fraction of its
lifetime.  Such a pre-existing void could result in a remnant which is
freely flowing to the north but has hit the edges to the south,
forming the visible, non-thermal shell.  Presuming that the remnant
has just recently reached the southern boundary of any pre-existing
void, we can assume that it has been expanding without hinderance for
most of its lifetime.  For an initial expansion velocity of $10^4$
\kms, this yields an age of only about 7000 yr, which we consider to
be an extreme lower limit.

The clumpy morphology and faintness of G55.0+0.3 leaves no doubt that
the true age lies between the limits described above.  Even though the
upper limit is quite high, such ages cannot be dismissed out of hand
as they are consistent with theoretical models.  Expansion into a
singular stellar wind bubble does not appear to occupy a significant
fraction of the lifetime of a radiative remnant.  The SNR is
significantly younger only if it has expanded into a much larger
cavity such as that which might be produced by a cluster of massive
stars.  The density estimate obtained from the H{\sc i} is too high to
suggest that an evacuated cavity exists now outside the continuum
shell.  Taking into account the likelihood that the remnant expanded
into a pre-existing cavity for at least part of its lifetime,
G55.0+0.3's radiative age of $\sim 10^6$ yr is taken to be an upper
limit, although we reiterate that its size and morphology are
indicative that it is likely in the radiative phase.

\subsection{An SNR/Pulsar Association?}

The proximity of PSR J1932+2020 to G55.0+0.3 begs the question as to
whether or not these objects are related.  Several factors argue in
favour of an association.

Using the position of PSR J1932+2020 as given in Table
\ref{psr-parameters} and a blast position of $19^{\rm h} 32^{\rm m}
4\fs 2$, $19^{\rm d}$ 58\arcmin \ 59\farcs 0 (the centre of the circle
in Figure \ref{channel93}), the pulsar's current position is $\sim
22$\arcmin \ from the assumed position of the SN which created
G55.0+0.3.  This means the pulsar lies at an angular distance of only
1.3 times the radius of the remnant.  To examine the possibility of a
physical association further, we consider estimates of the distance
and age of the pulsar.

From the dispersion measure, the distance of PSR J1932+2020 can be
estimated to be approximately 9 kpc, with an uncertainty of 25\%
(\markcite{tay93}Taylor \& Cordes 1993).  Measurements of H{\sc i}
absorption yield a lower limit of the turning point velocity to the
$v_{LSR}$ of the pulsar.  \markcite{fra91} Frail et al.~(1991) measure
an upper limit in H{\sc i} absorption of $-57$ \kms.  This is
consistent with our limit of about $-53$ \kms\ for the H{\sc i} void
we associate with G55.0+0.3.

The characteristic age (i.e.~spindown age $= P /2\dot{P}$) of pulsars
provides a lower limit on their true age (although its value
approaches the true value for large ages).  The characteristic age of
PSR J1932+2020 is $1.1 \times 10^6$ yr, based on data of
\markcite{lor95}Lorimer et al.~(1995).  This is consistent at the
2$\sigma$ level with our estimate of the upper limit of the age of
G55.0+0.3, $(1.9\pm 0.4) \times 10^6$ yr.

Using the pulsar's spindown age of 1.1 million years, its average
transverse velocity required to place it at the centre of the circular
fit to G55.0+0.3 at birth is 80 \kms.  The actual space velocity will
be larger than this, depending on the inclination of the pulsar's
velocity to the line of sight.

Proper motion measurements of \markcite{lyne82}Lyne, Anderson, \&
Salter (1982) have shown that pulsars are high-velocity objects.
Using 71 objects, \markcite{cor86}Cordes (1986) derived the mean
transverse speed of pulsars to be 100 \kms, with a tail to the
distribution extending to 300 \kms.  \markcite{lyn94}Lyne and Lorimer
(1994) have estimated the mean transverse pulsar birth speed to be
$345\pm 70$ \kms\ (rms 499 \kms) based on 29 pulsars younger than 3
Myr.  The ten oldest pulsars give a mean velocity of $105\pm 25$ \kms.
The inferred velocity of $\sim 80$ \kms \ is thus in good agreement with
typical old pulsar velocities.

The consistency of both the distance and age estimates of the pulsar
and the remnant suggests that the two could be associated and could
have been formed in the same SN event.  If this is indeed the case,
then the age of the SNR would be better established by that of the
pulsar, since spindown estimates become more accurate as pulsars age.

Most of the SNR/pulsar associations which have been proposed are for
the youngest known pulsars. Only one, CTB 80, has an age on the order
of $10^5$ yr, based on the spindown age of the pulsar B1951+32
(\markcite{shu89}Shull et al.~1989). Most are $\sim 20,000$ yr old
(\markcite{car95}Caraveo 1995).  This is not surprising, since most
have been found by focussed searches for associated SNR about young
pulsars.  This potential association of G55.0+0.3 and PSR J1932+2020
suggests that targeted searches should not be restricted by pulsars'
characteristic ages.

From a statistical analysis, \markcite{shu89}Shull et al.~(1989)
suggested that less than 1\% of SNRs last for $10^6$ yr, while fewer
than 20\% are still intact after $5 \times 10^5$ yr.  If rare SNRs of
ages on the order of a million years are visible, then the 20\% which
reach ages of $5 \times 10^5$ yr should also be detectable.  These
SNRs may simply have been too faint to be detected in previous radio
surveys.  Thus, the identification of such remnants is important for
the statistics of both SNR properties and SNR/pulsar associations.

The age estimate of SNR G55.0+0.3 is well above the average age of the
currently known SNR population of $\sim 60,000$ yr
(\markcite{fra94}Frail et al.~1994).  At greater than one million
years, the age of G55.0+0.3 is the oldest estimate for any radio
detected SNR.

\section{Conclusions}
\label{sec-conc}

Based on the analysis of available radio and infrared data, we have
determined that the composite object G55.2+0.5 contains one
non-thermal source, designated G55.0+0.3.  This radio half-shell has
an average spectral index of $-0.53\pm 0.26$ and shows distinctly
non-thermal emission across the majority of its surface.  The radio
spectrum of this object is consistent only with that of a shell-type
SNR.  The infrared/radio ratios for the region are also consistent
with the values previously observed for shell-type SNRs.  A spectral
index of $0.01\pm 0.10$ for G55.6+0.7 to the north and the
heightened far-infrared emission are consistent with the results of
\markcite{rei86}Reich et al.~(1986), who concluded that the object is
an H{\sc ii} region.

H{\sc i} line data show a void coincident with the continuum
emission at a kinematic distance of 14 kpc.  This implies a radius of
70 pc for the SNR at a height of $\sim 73$ pc above the Galactic plane.
Analysis of the H{\sc i} emission surrounding the void yields an
ambient density of $0.8 \pm 0.2$ cm$^{-3}$.  If the remnant is
expanding into this material, it is in the radiative phase, and its age
is estimated to be $(1.9\pm 0.4) \times 10^6$ yr.

The assertion that the remnant is associated with the nearby pulsar,
J1932+2020, is intriguing.  The remnant and pulsar have comparable
estimates of age and distance.  If valid, the association would
confirm the age of the SNR.  At 14 kpc, a transverse velocity of 80
\kms \ corresponds to a proper motion of 1-2 mas per year.  A proper
motion measurement of the pulsar would aid in confirming or disproving
the proposed association.

G55.0+0.3 is among the faintest SNRs known.  This SNR could be just
one member of a larger population of faint, old remnants that are not
currently detectable at radio frequencies.  If a significant fraction
of SNR survive longer than $5 \times 10^5$ yr, further imaging of the
Galactic plane with high surface brightness sensitivity and high
angular resolution should reveal more old SNRs.

\vspace{40pt}

This research was supported by a grant to ART from the Natural
Sciences and Engineering Research Council of Canada.  The DRAO
Synthesis Telescope is operated as a national facility by the National
Research Council of Canada.  BCM would like to thank ART and BJW for
insightful feedback and strong encouragement throughout this project.
We would also like to thank our referee for an insightful and thorough
review.

\clearpage

\figcaption[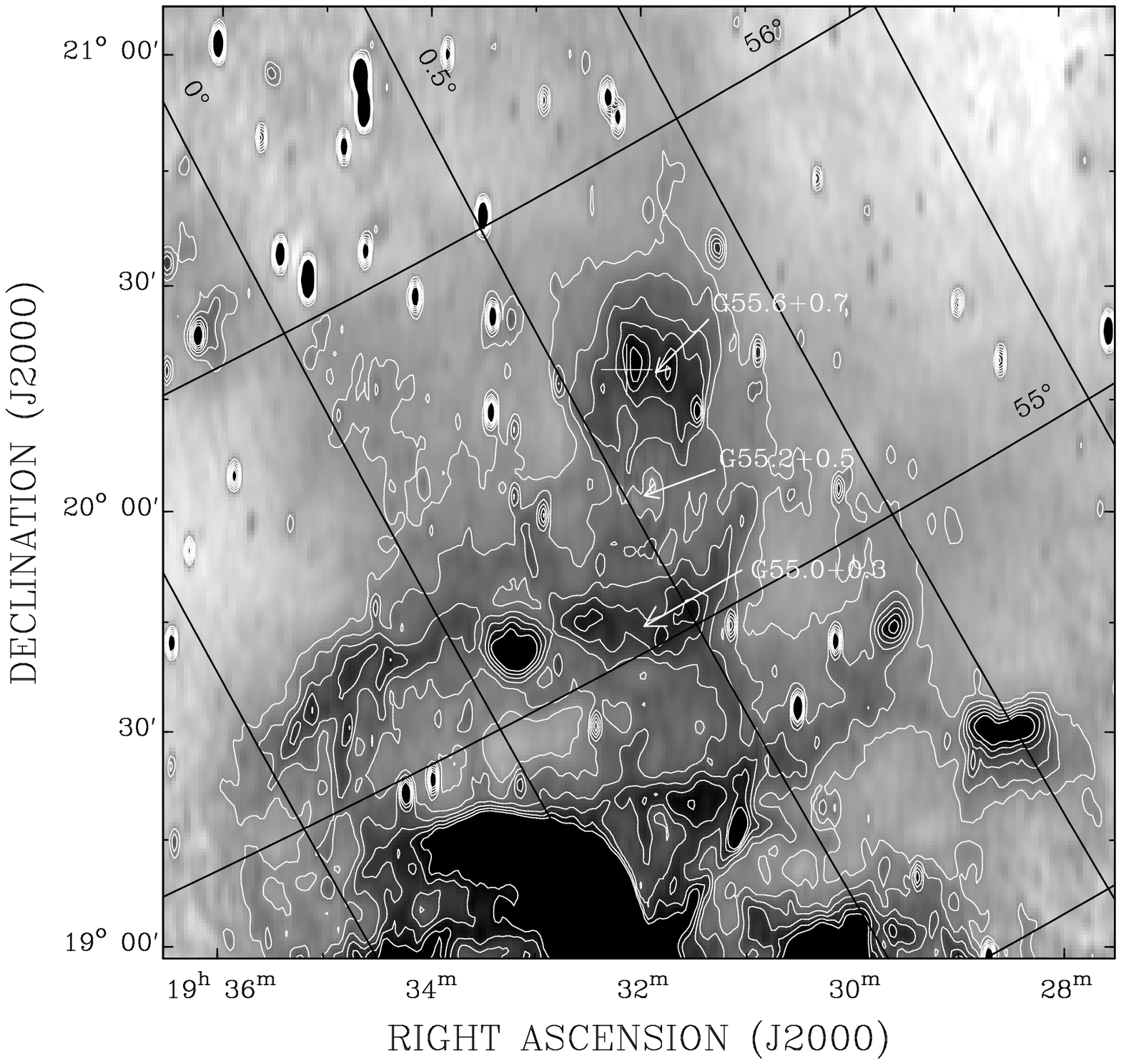]{These data are the 327 (left)
and 1420 (right) MHz images.  The images are 256 $\times$ 256 pixels
with a grid scale of 30\arcsec.  The greyscale limits on the 327 MHz
map are $-30$ (white) and 40 K(T$_b$) (black).  The contour values are
10, 15, 20 and 30 K(T$_b$).  The limits of the 1420 MHz map are 9.5
(white) and 13 K(T$_b$) (black), with contours ranging from 11 to 13
K(T$_b$) in increments of 0.4 K(T$_b$).  The resolution of both maps
is 1.0\arcmin \ $\times$ 2.92\arcmin. \label{cont-images}}

\figcaption[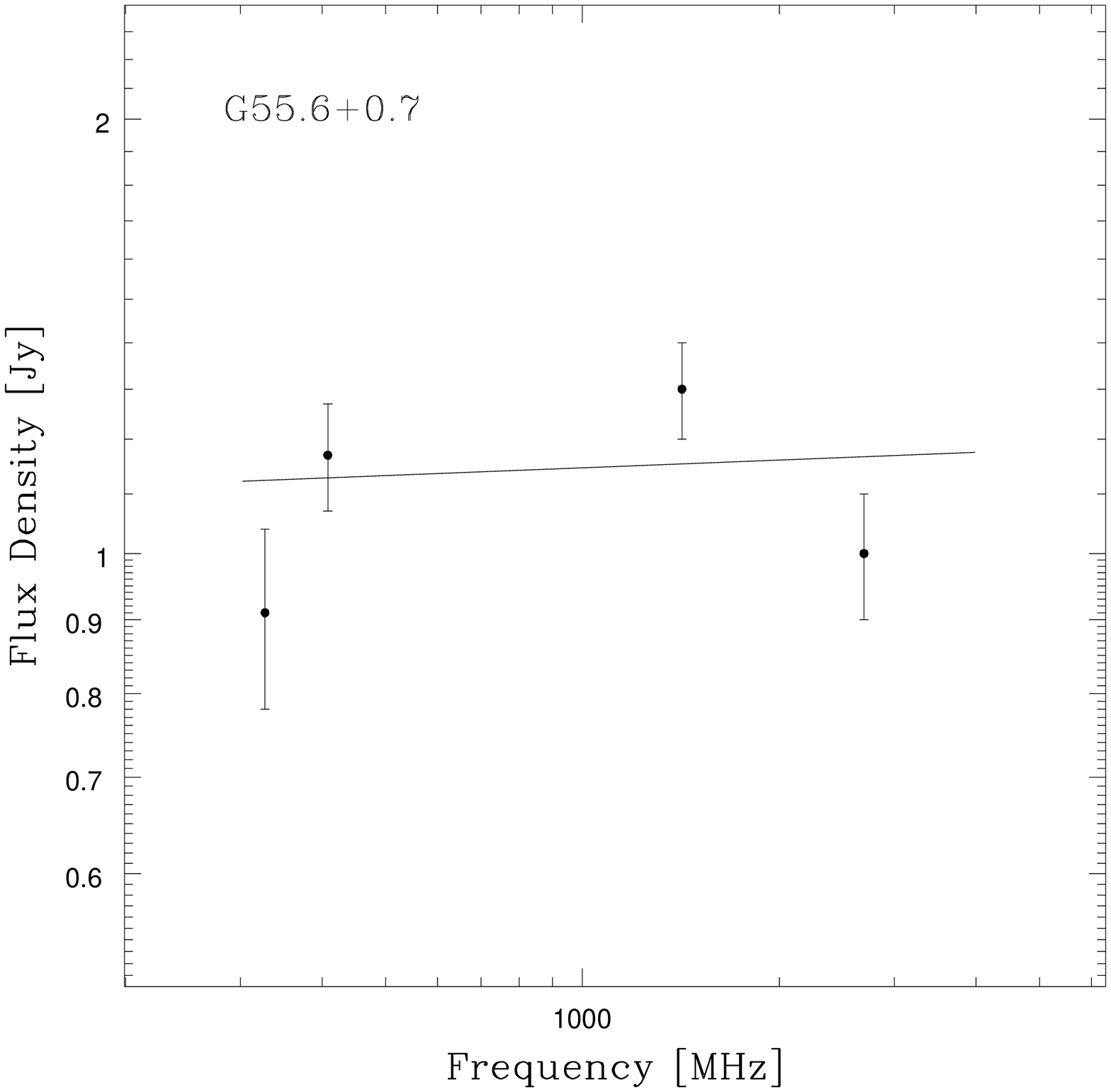]{The radio spectrum of G55.6+0.7 was
generated with four frequencies (327, 408, 1420 and 2695 MHz) and is
nearly flat.  The spectrum for G55.0+0.3 (calculated without the flux
density at 408 MHz due to confusion) is non-thermal.  The flux
densities at 327 MHz are underestimates due to the absence of
short-spacings.  The spectra were calculated after the effects of the
pulsar and point sources were removed from each frequency. The error
bars indicate 1$\sigma$ standard deviations from the least squares
fits.  The slopes result in spectral indices of $0.01\pm 0.10$ and
$-0.53\pm 0.26$, respectively. \label{spectra}}

\figcaption[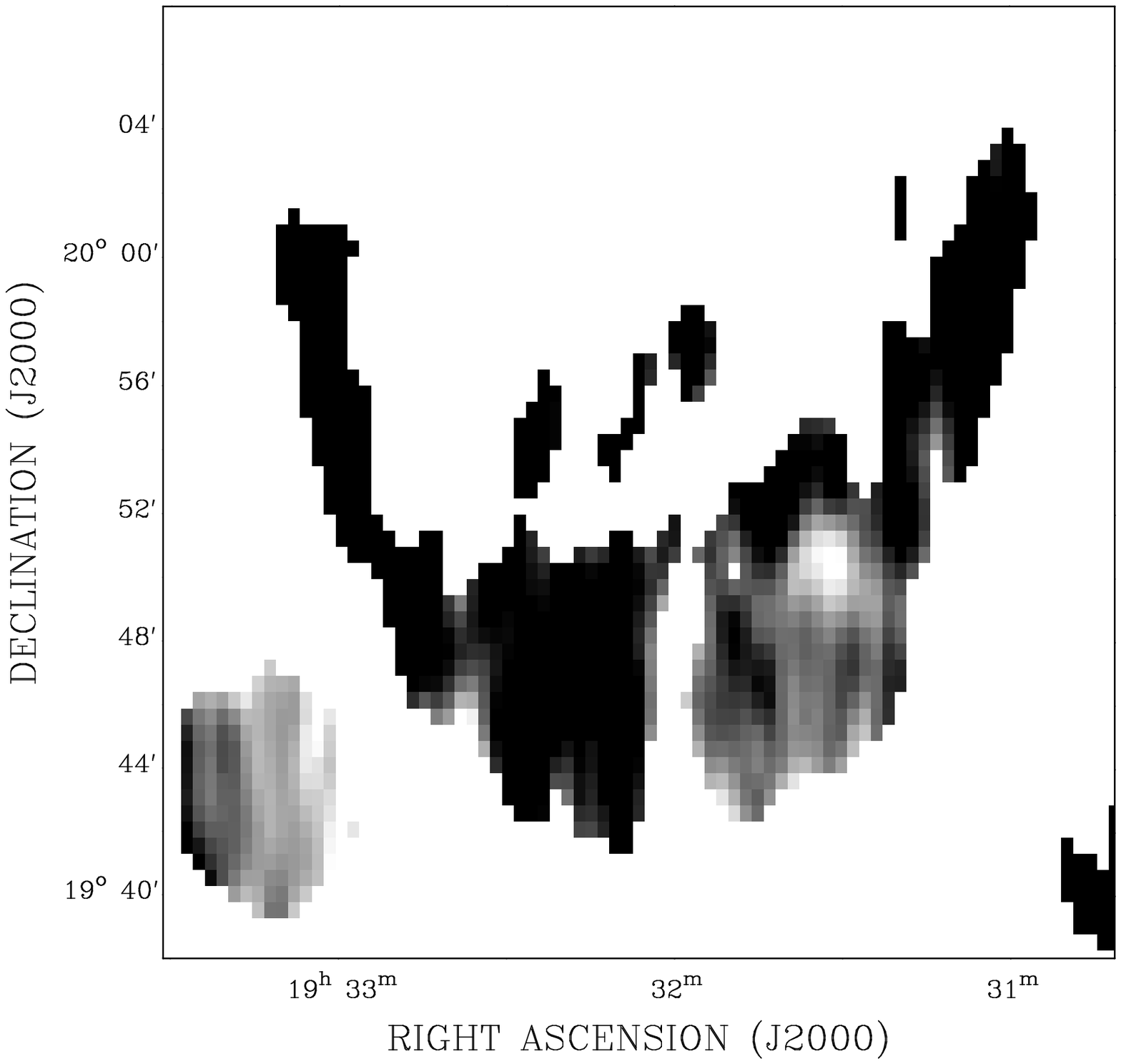]{This spectral index map of
G55.0+0.3 shows that not all of the shell revealed a non-thermal
spectral index.  Brightness temperature spectral indices less than 1.7
(or undefined on the exterior) appear white, while those greater than
2.2 (non-thermal) are black.  Here, the spectral index is defined by
$T_b \propto \nu^{2-\alpha}$, alpha given by $S_\nu \propto
\nu^\alpha$.  One large section in the southwest shows a spectral
index that could be thermal.  The circular region to the southeast of
the SNR is most probably an H{\sc ii} region.  \label{mouth}}

\figcaption[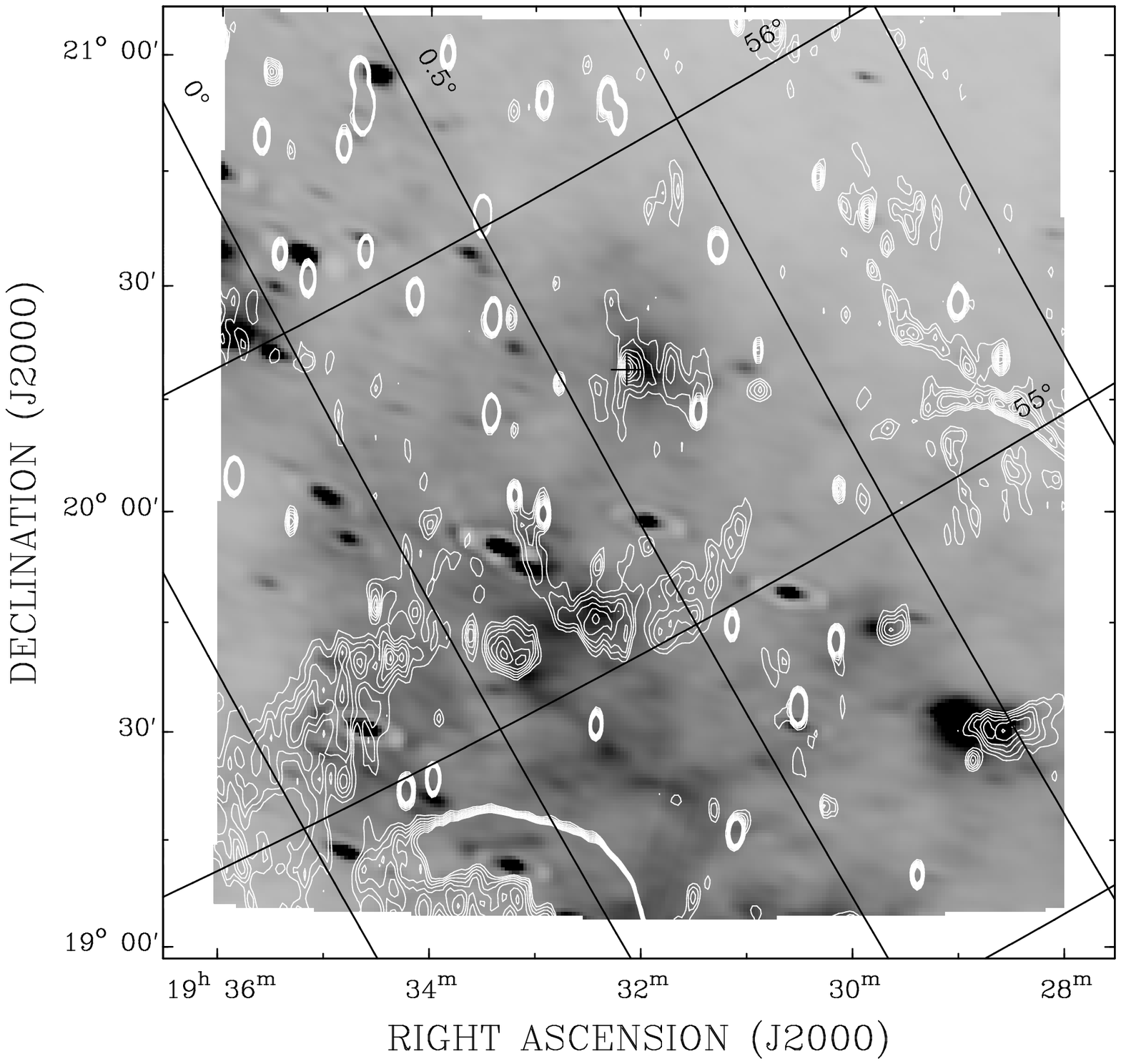]{This figure compares
infrared to radio emission over the same region as the maps of Figure
\ref{cont-images}.  The greyscale shows the distribution of infrared
emission at 60 $\mu$m.  Black indicates emission greater than 150 MJy
sr$^{-1}$, while white indicates emission below zero.  The overlaid
contours at 327 MHz have values of 10, 15, 20, 25, 30, 35, 40 and 45
K($T_b$).  Note that there is a source of IRAS emission coincident
with the radio source G55.6+0.7; however, the southern shell has no
such counterpart, containing only a semi-compact infrared source near
its southeast corner.  Some infrared point sources were subtracted
from the map for the determination of flux densities using polygons.
The black cross marks the position of the pulsar J1932+2020.
\label{iras-radio}}

\figcaption[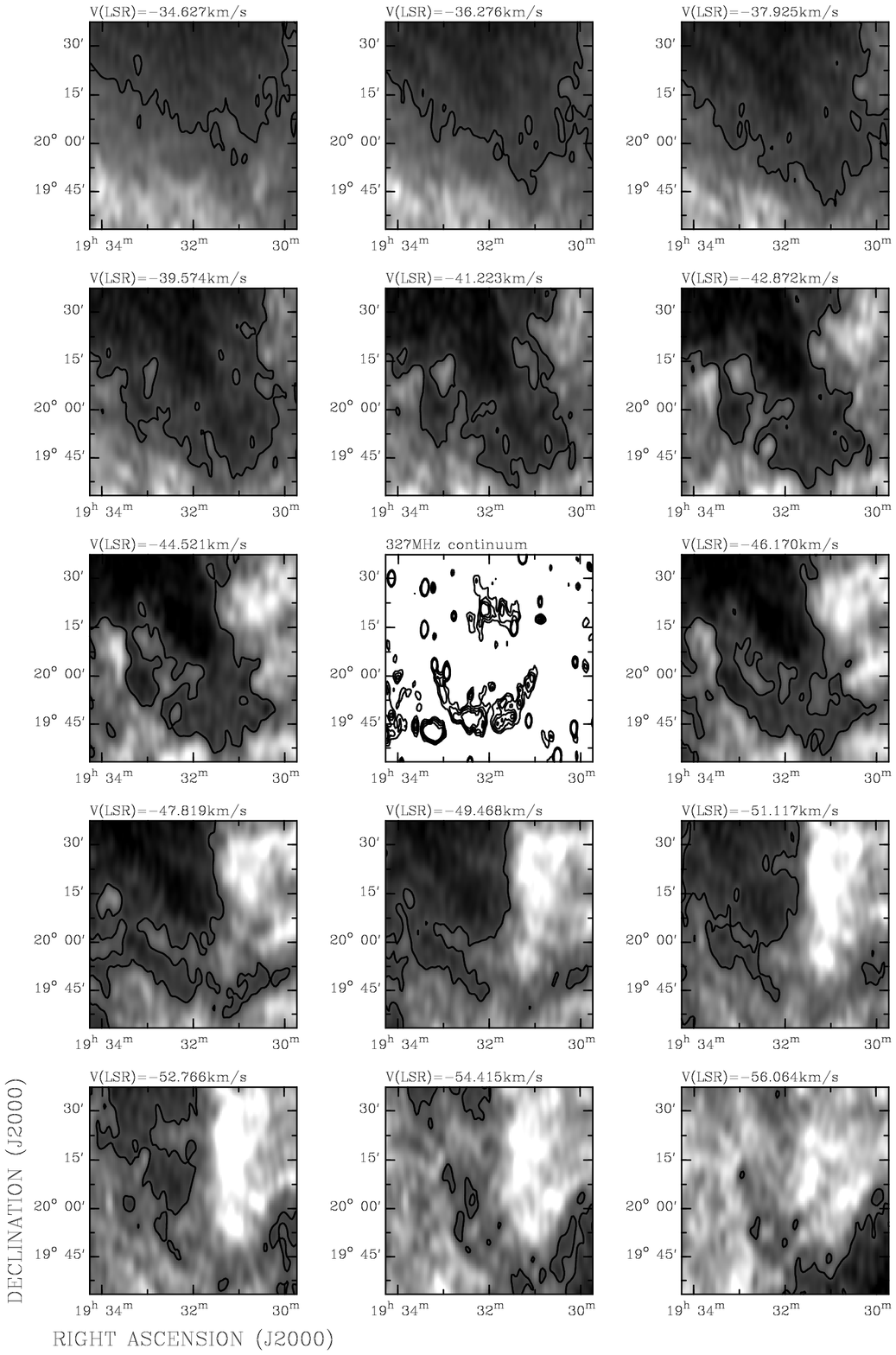]{H{\sc i} Velocity
Channels: $-34.6$ to $-56.1$ \kms.  These bias-subtracted maps have
limits of $-15$ (black) and 20 (white) K($T_b$).  A single black
contour is included at $-5$ K($T_b$).  A 327 MHz continuum contour map
is shown in the centre with contours of 10, 15, 20 and 25 K($T_b$).
The continuum map is included separately as its contours tend to
obscure the H{\sc i} below.  Bias-subtraction was performed solely to
allow the presentation of all maps on the same scale.  A larger image
of the central channel is shown in Figure \ref{channel93}.  A void
(partial or total) coincident with the continuum shell can be seen
from velocities of $-39.6$ to $-52.8$ \kms.  Several channels, where
no structures correlated to the continuum shell were seen, are shown
on either side of these.
\label{HI89-97fig}}

\figcaption[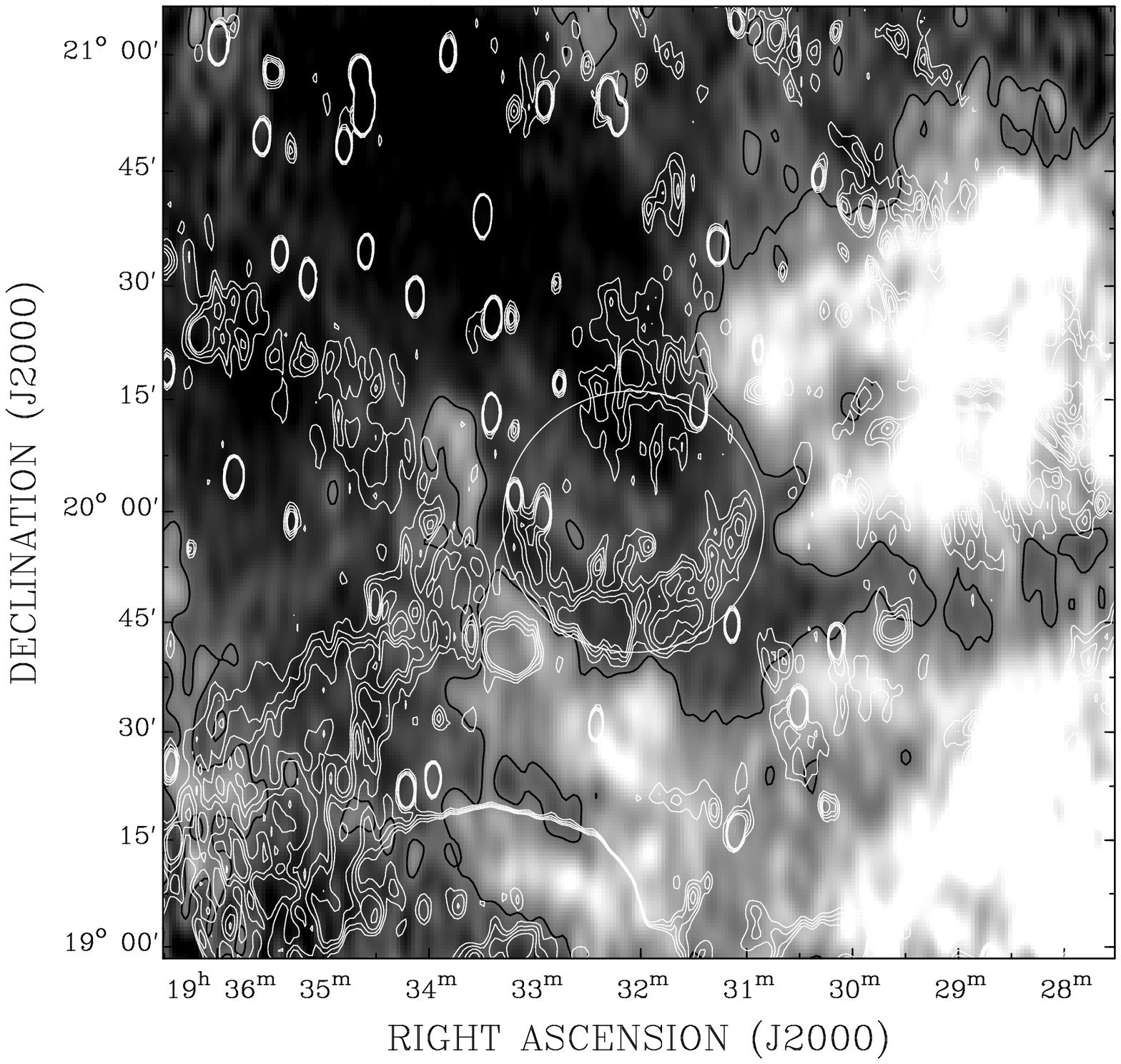]{Central H{\sc i} map for
$v_{LSR} = -46.2$ \kms.  The shell at 327 MHz in white contours (5,
10, 15 and 20 K($T_b$)) is matched well by the H{\sc i} greyscale void
at $-46.2$ \kms.  Many of the variations in the continuum are echoed
in the void, especially to the southeast.  The greyscale limits for
the H{\sc i} are 20 (white) and $-15$ K($T_b$) (black).  The single
black contour in H{\sc i} is placed at 0 K($T_b$). The map has been
bias-subtracted.  The 35\arcmin \ circle indicates an approximate diameter
for the remnant. \label{channel93}}

\figcaption[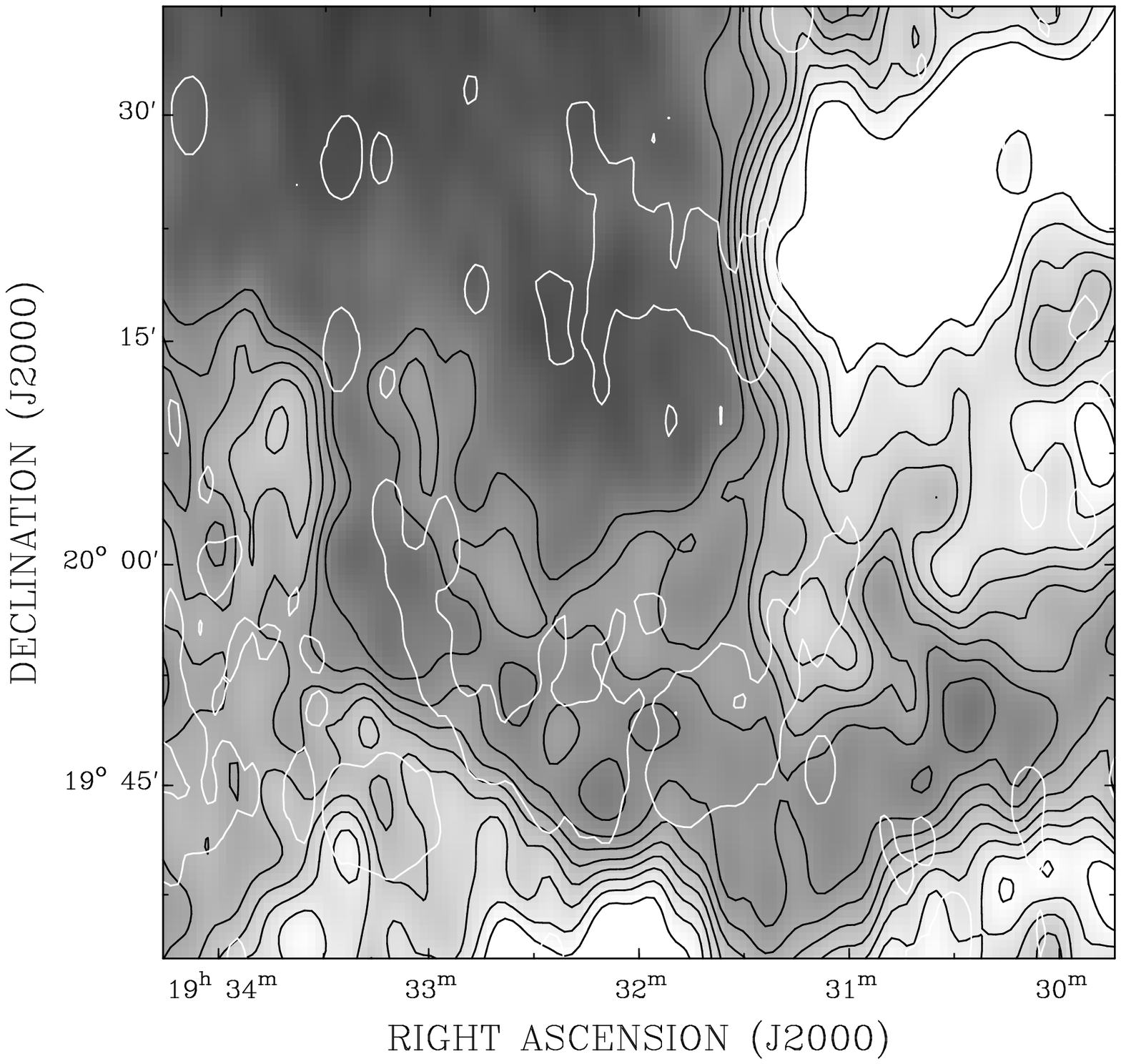]{This density map was calculated under
the assumptions of spherical expansion of the remnant.  The map has a
grid spacing of 30\arcsec.  The greyscale limits are 0 (black) to 3
cm$^{-3}$ (white), while the contours are 1.4, 1.6, 1.8 and 2
cm$^{-3}$ respectively. A single white contour at 10 K($T_b$) of the
327 MHz continuum has been added to indicate the location of the
shell. The estimated uncertainty on each density measurement is 6\%.
\label{density}}

\clearpage

\begin{table}
\caption{\bf DRAO Synthesis Data Specifications}
\begin{small}
\begin{tabular}{lcc} \hline \hline
& 408 MHz ($\lambda $74 cm) & 1420 MHz ($\lambda $21 cm) \\ \hline
central frequency & 408 MHz & 1420.406 MHz \\
polarization & left circular & right circular \\
continuum bandwidth & 4 MHz & 30 MHz \\
system temperatures & 150 K & 65 K \\
field size (to 20\% response) & 8.1\arcdeg & 2.6\arcdeg \\
synthesized beam (EW $\times$ NS) & 3.4\arcmin \ $\times$ 9.8\arcmin & 
1.0\arcmin \ $\times$ 2.92\arcmin \\
measured rms noise ($T_b$) & 854 mK & 40 mK \\ \hline \hline
\end{tabular}
\end{small}
\label{DRAOspecs-st}
\end{table}

\clearpage

\begin{table}
\caption{\bf Specifications of Synthesis H{\sc i} Spectral Line Data}
\begin{small}
\begin{tabular}{ll} \hline \hline
polarization & left and right circular\\
bandwidth & 1 MHz \\
field size (to 20\% response) & 2.6\arcdeg \\
synthesized beam (EW x NS) & 1.74\arcmin \ $\times$ 5.10\arcmin \\
rms noise in the map centre/channel ($T_b$) & 1.17 K \\
central velocity & 0 \kms \\
total 128 channel span & 211 \kms \\
channel separation & $\sim 1.65$ \kms \\
channel width & $\sim 2.6$ \kms \\ \hline \hline
\end{tabular}
\end{small}
\label{DRAOspecs-HIst}
\end{table}

\clearpage

\begin{table}
\caption{\bf Specifications of Images at Alternate Wavelengths}
\begin{small}
\begin{tabular}{lccc} \hline \hline
$\lambda$/$\nu$ & Resolution & PA (N of E) & rms noise \\
\hline
60 $\mu$m & 1.67\arcmin \ $\times$ 1\arcmin & 340\arcdeg & 
21\%\tablenotemark{a} \\
100 $\mu$m & 2.42\arcmin \ $\times$ 1.92\arcmin & 335\arcdeg & 
32\%\tablenotemark{a} \\
11 cm / 2695 MHz & 4.27\arcmin \ $\times$ 4.27\arcmin & --- & 24 mK($T_b$) \\
92 cm / 327 MHz & 1\arcmin \ $\times$ 2.92\arcmin & 90\arcdeg & 4 K($T_b$) \\
\hline \hline 
\end{tabular}
\tablenotetext{a}{Infrared Processing and Analysis Center estimated
uncertainty in spatially integrated flux densities over a source (Fich
\& Terebey 1996)}
\end{small}
\label{otherfreqtable}
\end{table}

\clearpage

\begin{table}
\caption{\bf Established Parameters of PSR J1932+2020}
\begin{small}
\begin{tabular}{lcc} \hline \hline
& & 1$\sigma$  \\
RA & $19^{\rm h} 32^{\rm m} 8\fs 30$ & 3 \\
Dec & $20^{\rm d}$ 20\arcmin \ 46\farcs 30 & 6 \\
Period, $P$ & 0.268216854361 s & 3 \\
$\dot{P} (\times 10^{-15}) $ & 4.21630 & 18 \\
dispersion measure & 211.007 cm$^{-3}$pc & 17 \\ \hline \hline
\end{tabular}
\end{small}
\label{psr-parameters}
\end{table}

\clearpage

\begin{table}
\caption{\bf Integrated Flux Densities over a Series of Polygons}
\begin{small}
\begin{tabular}{lccccc} \hline \hline
Region & \# polygons & $S_{2695}$ & $S_{1420}$ & $ S_{408}$ &
$S_{327}$ \\ 
& averaged & (Jy) & (Jy) & (Jy) & (Jy) \\ \hline 
east G55.6+0.7 & 13 & --- & $0.088 \pm 0.012$ & --- & $0.057 \pm 0.013 $ \\
west G55.6+0.7 & 11 & --- & $0.037 \pm 0.005$ & --- & $0.038 \pm 0.008 $ \\ 
G55.6+0.7 & 7 & $ 1.0 \pm 0.1$ & $1.3\pm 0.1$ &
$1.17 \pm 0.14$ & $0.91 \pm 0.13$ \\ 
G55.0+0.3 & 14 & $ 0.27 \pm 0.05$ & $0.64 \pm 0.11$ & $0.25 \pm 0.12$
& $0.98 \pm 0.15 $ \\ 
G55.2+0.5 & 12 & $ 1.7 \pm 0.1$ &
$2.9 \pm 0.3$ & $2.4 \pm 0.3$ & $2.75 \pm 0.31 $ \\ \hline \hline
\end{tabular}
\end{small}
\label{intfluxtable}
\end{table}

\clearpage

\begin{table}
\caption{\bf Radio Spectral Indices and Infrared/Radio Ratios}
\begin{small}
\begin{tabular}{lccccc} \hline \hline
\hspace{2mm}& east G55.6+0.7 & west G55.6+0.7 & G55.6+0.7 & G55.0+0.3 &
G55.2+0.5 \\ \hline
\hspace{1cm}$\alpha$\hspace{8mm} & $ 0.30 \pm 0.13 $ & $ -0.01 \pm 0.10 $ & 
$0.01 \pm 0.10 $ & $ -0.53 \pm 0.26 $ & $ -0.21 \pm 0.19 $ \\ 
$S_{60 \mu m}/S_{327 \rm{MHz}}$ & $1063\pm 361 $ & $237\pm 155 $ & 
$503\pm 121 $ & $155\pm 53 $ & $275\pm 60 $ \\ \hline \hline
\end{tabular}
\end{small}
\label{intalphatable}
\end{table}

\clearpage

\begin{table}
\caption{\bf SNRs of Large Radii}
\begin{small}
\begin{tabular}{lcccl} \hline \hline
SNR & Angular Size & $d_l$ (kpc) & $R_{avg}$ (pc) & Reference \\
\hline 
G160.9+2.6 (HB 9) & 140\arcmin \ $\times$ 120\arcmin & $<$4 & $<$150 &
\markcite{lea91}Leahy \& Roger (1991) \\ 
G39.7$-2.0$ & 120\arcmin \ $\times$ 60\arcmin & 4.5 & 120 & 
\markcite{rom87}Romney et al.~(1987) \\
G166.2+2.5 (OA 184) & 90\arcmin \ $\times$ 70\arcmin & 8 & 90 & 
\markcite{rou86}Routledge et al.~(1986) \\
G116.5+1.1 & 80\arcmin \ $\times$ 60\arcmin & 3.6 & 70 & 
\markcite{rei81}Reich et al.~(1981) \\ 
G114.3+0.3 & 90\arcmin \ $\times$ 55\arcmin & 3 & 60 &
\markcite{fur93}F\"{u}rst et al.~(1993) \\ 
G205.5+0.5 & 220\arcmin \ $\times$ 220\arcmin & 0.8 & 50 & 
\markcite{gra82}Graham et al.~(1982) \\ \hline \hline
\end{tabular}
\end{small}
\label{sizes}
\end{table}








\end{document}